\documentclass[%
 reprint,
 amsmath,amssymb,
 aps,
 pre
]{revtex4-2}

\usepackage[english]{babel}
\usepackage{amssymb}
\usepackage[utf8]{inputenc}
\usepackage[T1]{fontenc}
\usepackage{tabularx}
\usepackage{siunitx}
\usepackage{comment}

\usepackage{float}
\usepackage{enumitem}
\usepackage{mathrsfs}

\addto\captionsenglish{}

\usepackage{amsmath}
\usepackage{mathtools}
\usepackage{graphicx}
\usepackage[colorinlistoftodos]{todonotes}
\usepackage[colorlinks=true, allcolors=blue]{hyperref}
\usepackage[titletoc,toc,title]{appendix}
\usepackage{algorithm}
\usepackage[noend]{algpseudocode}

\usepackage{csquotes}

\begin{document}

\title{Dynamics of Time-Modulated, Nonlinear Phononic Lattices}

\author{B.~L.~Kim}
\author{C.~Daraio}
\email{daraio@caltech.edu}
\affiliation{Department of Mechanical and Civil Engineering, California Institute of Technology, Pasadena, CA 91125, USA}
\author{C.~Chong}
\affiliation{Department of Mathematics, Bowdoin College, Brunswick, ME 04011, USA}
\author{S.~Hajarolasvadi}
\altaffiliation[Also at ]{Department of Mechanical and Civil Engineering, California Institute of Technology}
\affiliation{Department of Civil and Environmental Engineering, University of Illinois at Urbana-Champaign, Urbana, IL 61801, USA}
\author{Y.~Wang}
\affiliation{School of Mechanical and Aerospace Engineering, Nanyang Technological University, Singapore, Singapore 6397984}

\date{\today}

\begin{abstract}
The propagation of acoustic and elastic waves in time-varying, spatially homogeneous media can exhibit different phenomena when compared to traditional spatially-varying, temporally-homogeneous media. In the present work, the response of a one-dimensional phononic lattice with time-periodic elastic properties is studied with experimental, numerical and theoretical approaches. The system consists of repelling magnetic masses with grounding stiffness controlled by electrical coils driven with electrical signals that vary periodically in time. For small amplitude excitation, in agreement with theoretical predictions, wavenumber bandgaps emerge. The underlying instabilities associated to the wavenumber bandgaps are investigated with Floquet theory and the resulting parametric amplification is observed in both theory and experiments. In contrast to genuinely linear systems, large amplitude responses are stabilized via the nonlinear nature of the magnetic interactions of the system. In particular, the parametric amplification induced by the wavenumber bandgap can lead to bounded and stable responses that are temporally quasi-periodic. Controlling the propagation of acoustic and elastic waves by balancing nonlinearity and external modulation offers a new dimension in the realization of advanced signal processing and telecommunication devices. For example, it could enable time-varying, cross-frequency operation, mode- and frequency-conversion, and signal-to-noise ratio enhancements. 

\end{abstract}

\maketitle


\section{Introduction}
Acoustic metamaterials and phononic crystals often achieve control of wave propagation by leveraging scattering effects induced by the presence of spatial periodicity in the design of their micro-structure \cite{martinez-sala_sound_1995,james_sonic_1995,kushwaha_acoustic_1993,vasseur_experimental_2001,lu_phononic_2009}. In active mechanical systems, the periodic variation of material properties in time provides an additional, less-explored strategy to control acoustic and elastic waves. This strategy draws inspiration from the study of parametric amplifiers \cite{cullen_travelling-wave_1958,tien_traveling-wave_1958} and the effects of traveling-wave-like harmonic modulations (i.e., periodic in both space and time) on electromagnetic waves \cite{cassedy_dispersion_1963,cassedy_dispersion_1967}. Indeed, spatio-temporally periodic acoustic and elastic systems have been shown, both theoretically and experimentally, to exhibit the characteristic opening of bandgaps in their dispersion relations. Most studies have focused on the presence of nonreciprocal frequency bandgaps in linear systems, such as beams and metamaterials with spatially and temporally varying resonators or continuous elastic structures with spatio-temporally periodic properties \cite{trainiti_non-reciprocal_2016,nassar_modulated_2017,nassar_non-reciprocal_2017,nassar_non-reciprocal_2017-1,ansari_application_2017,wang_observation_2018,goldsberry_non-reciprocal_2019,chen_nonreciprocal_2019,zhu_non-reciprocal_2020,marconi_experimental_2020,nassar_nonreciprocity_2020}.

Whereas frequency bandgaps are the hallmark of spatially periodic systems, gaps in the wavenumber axis of a linear medium’s dispersion relation have been shown to arise in time-periodic and spatio-temporally periodic systems, in which the wave speed of the traveling-wave-like-modulation is greater than the velocity of propagation of the medium \cite{cassedy_dispersion_1967, reyes-ayona_observation_2015, wang_observation_2018, galiffi_broadband_2019, trainiti_time-periodic_2019, lee_parametric_2021}. These so-called wavenumber bandgaps are understood to be parametrically amplified standing waves (i.e., non-propagating, hence the analogous notion of a bandgap) \cite{cullen_travelling-wave_1958,reyes-ayona_observation_2015,trainiti_time-periodic_2019}. For example, in a phononic lattice with a supersonic traveling-wave modulation, incident signals within the induced wavenumber bandgap excite unstable oscillations of the scattered field. This results in apparent amplification of frequencies corresponding to the bandgaps, which are different for forward- and backward-propagating waves since this form of spatio-temporal modulation breaks reciprocity \cite{wang_observation_2018}. 
In another example, an elastic waveguide is modulated periodically only in time via an array of piezoelectric patches controlling the stiffness. The reflection of a broadband signal incident on the interface of the modulated region is observed to comprise narrowband content at half the modulation frequency, consistent with the parametrically amplified standing wave solution present within wavenumber bandgaps \cite{trainiti_time-periodic_2019}. Wavenumber bandgaps have been shown  to open experimentally in transmission lines and theoretically in proposed photonic systems \cite{reyes-ayona_observation_2015,lee_parametric_2021}. In a system more analogous to the present study, instabilities in a linear phononic lattice with time-modulated spatially periodic modulations have been explored \cite{li_wave_2014}, but the opening of wavenumber bandgaps in the dispersion relation of phononic systems has not been directly shown.

Aside from periodicity, the introduction of nonlinearity provides an alternative strategy to control waves in discrete chains. The role of nonlinearity in discrete chains, for example, has been studied extensively since the first analysis of the Fermi-Pasta–Ulam–Tsingou problem \cite{fermi_studies_1955,ford_fermi-pasta-ulam_1992}. Some examples include nonlinearity-induced bandgaps \cite{chen_study_2007,boechler_tunable_2011,gantzounis_granular_2013}, nonreciprocal transmission \cite{boechler_bifurcation-based_2011}, discrete breathers \cite{daraio_energy_2006,boechler_discrete_2010,theocharis_intrinsic_2010}, solitary waves \cite{toda_wave_1967,nesterenko_propagation_1983,moleron_solitary_2014} frequency conversion \cite{serra-garcia_tunable_2018}, and nonlinear dispersion \cite{narisetti_perturbation_2010,manktelow_weakly_2014,wang_influences_2016,fang_wave_2017}. A more comprehensive review of the extensive work done on nonlinear lattices can be found in the review articles \cite{kevrekidis_non-linear_2011,berman_fermi-pasta-ulam_2005,flach_discrete_2008} or books \cite{starosvetsky_dynamics_2017,kevrekidis_discrete_2009,chong_coherent_2018,nesterenko_dynamics_2001}.

Nonlinear effects and their interplay with parametric amplification have indeed been studied in photonic and transmission line systems, which serve as practically implementable analogs to one-dimensional optical metamaterials. 
Multistability has been shown in Kerr nonlinear photonic crystals \cite{soljacic_optimal_2002,wang_optical_2008,wen_tunable_2020} as well as transmission lines with nonlinear capacitance \cite{powell_multistability_2008}. Parametric amplification in nonlinear transmission lines has also been demonstrated \cite{kozyrev_parametric_2006,powell_asymmetric_2009}.
Unidirectional soliton-like edge states in nonlinear Floquet topological insulators, which are modeled
by a discrete Nonlinear Schr\"odinger equation with time variable coefficients, were explored in
\cite{mukherjee_observation_2021}.  The interactions between extrinsic time-periodic modulation and nonlinear effects in phononic systems, however, are far less explored.

In the present study, we investigate a time-modulated phononic lattice in both linear and nonlinear regimes using a combination of experiments, theory and numerical simulations.  The paper is structured as follows: The experimental platform and corresponding model equations
are detailed in Sections~\ref{sec:exp} and \ref{subsec:latticeDynamics}, respectively. Results on wavenumber bandgaps and their associated instabilities in the small amplitude regime and a parametric investigation of stability are reported in Section~\ref{sec:wnbg}.  Nonlinear effects leading to the formation of stable temporally quasi-periodic orbits are explored in \ref{sec:nonlinear}. Section~\ref{sec:theend} concludes the paper.


\section{Experimental Platform} \label{sec:exp}
\begin{figure}
    \includegraphics[width=\linewidth]{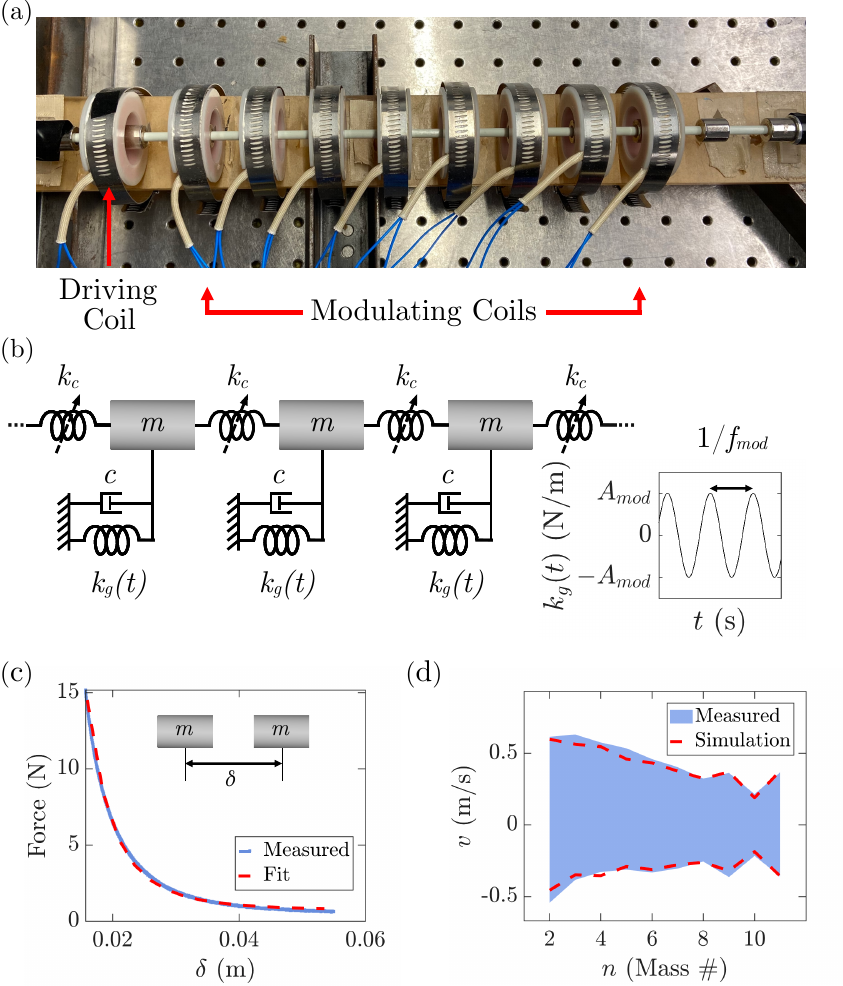}\\
    \caption{(Color online) \textbf{(a)} Photo of the experimental apparatus, with electromagnetic coils corresponding to grounding springs (modulating coils) and ring magnets inside each coil (not fully visible) sliding freely on a low-friction rod. \textbf{(b)} Schematic of the mass-spring lattice model with nonlinear coupling stiffness $k_c$, time-varying grounding stiffness $k_{g}(t)$, and viscous damping $c$; right panel shows the harmonic modulation form of the time-varying grounding stiffness. \textbf{(c)} Force-distance measurement (blue line) \cite{wang_observation_2018} and fit (red dashed line) of repulsive magnetic force between neighboring masses \textbf{(d)} Measured (blue shaded region) and numerically simulated (red dashed lines) nodal velocity envelope, used for viscous damping parameter fitting by matching spatial decay.}
    \label{fig:schematic_photo}
\end{figure}

The experimental setup is adapted from the platform developed by Wang et al. \cite{wang_observation_2018}. A one-dimensional (1D) phononic lattice is realized as a mass-spring chain composed of $N$ ring magnets (K\&J Magnetic, Inc., P/N R848) lined with sleeve bearings (McMaster-Carr P/N 6377K2) comprising the uniform masses, arranged with alternating polarity on a smooth rod (McMaster-Carr P/N 8543K28). Electromagnetic coils (APW Company SKU: FC-6489) are fixed concentrically around the equilibrium positions of each of the innermost eight masses, such that they may exert a restoring force on each mass proportional to an applied current. The chain has fixed boundary conditions, and the input mass (the first free inward mass from one of the fixed ends) has a concentric electromagnetic coil offset axially from its equilibrium position, which provides the driving force. Fig.~\ref{fig:schematic_photo}(a) shows the experimental setup. A lattice of size $N = 12$ (with fixed boundaries $n=1$ and $n=12$) is selected, to cover a sufficient range of wavelengths for the investigation of transmission and dispersion of waves, while minimizing the role of frictional losses.

The modulated magnetic lattice is modeled as a monatomic, nonlinear mass-spring chain with dynamically variable grounding springs and linear damping (see Fig.~\ref{fig:schematic_photo}(b)). The magnetic repulsive force between adjacent masses provides the nonlinear coupling stiffness, where the experimental force-distance relation is fit as a dipole approximation, as shown in Fig.~\ref{fig:schematic_photo}(c). For small displacements, the coupling stiffness may be approximated as a linear spring, taking the slope of the force-distance relation at the equilibrium mass spacing.
The modulating electromagnetic coils are modeled as a grounding stiffness applied to each mass, which may be positive or negative. A sinusoidal current applied uniformly (and in phase) to every modulating coil provides time-periodic grounding stiffness modulation $k_{g}(t)$, shown in Fig.~\ref{fig:schematic_photo}(b). Experimentally measured values for the force exerted on each mass by the magnetic field induced by each concentric electrical coil are used in analytical and numerical models (data from \cite{wang_observation_2018}). Dissipative forces, modeled as linear damping forces, are estimated as a fitting parameter (see Section \ref{subsec:latticeDynamics}).

\section{Model Equations}\label{subsec:latticeDynamics}

The lattice is modeled as a discrete mass-spring chain, wherein the equation of motion for the $n^{\mathrm{th}}$ mass (ring magnet) with displacement $u_n$ may be written as \cite{wang_observation_2018}
\begin{eqnarray} \label{eq:EOM}
    m\frac{d ^{2}{u}_{n}}{d t^{2}} + k_{g} (t) u_{n} + F_{loss,n} + F_{mag,n} =\\
    \delta_{2,n} A_{dr} \cos{\left(2\pi f_{dr} t \right),} \nonumber
\end{eqnarray}
for $n=1$ to $N$, with fixed boundary conditions $u_{1}(t) = u_{N}(t)=0$. All ring magnets have uniform mass $m$.
The variable grounding stiffness, $k_{g}(t)$, acts uniformly on every individual mass. The dissipative forces are represented by $F_{loss,n}$, and $F_{mag,n}$ is the coupling force acting on the $n^{\mathrm{th}}$ mass due to repulsive force between neighboring ring magnets. The driving input amplitude and frequency are given by $A_{dr} ([A_{dr}]=\text{N})$ and $f_{dr} ([f_{dr}] = 
\text{Hz})$, respectively. The Kronecker delta, $\delta_{2,n}$, acts so that the input forcing applies only to mass $n=2$.

An AC sinusoidal voltage (with zero DC offset) applied to the electromagnetic coils induces a harmonic grounding stiffness modulation of the form
\begin{equation}\label{eq:modulation}
    k_{g}(t) = \delta_{j,n} A_{mod} \cos{\left(2\pi f_{mod} t \right)}
\end{equation}
with amplitude $A_{mod} \;([A_{mod}]=\text{N/m})$ and frequency $f_{mod} \,([f_{mod}]=\text{Hz}) $ (see Fig.~\ref{fig:schematic_photo} (b) inset). The Kronecker delta index $j = 3$ to $N-2$ so that the modulation acts only on masses $n = 3$ to $N-2$, labeled as in Fig.~\ref{fig:schematic_photo}(a). The amplitudes $A_{dr}$ and $A_{mod}$ are determined using empirical relations between applied voltage and resultant current in the coils and measurements of the reaction force exerted by the coils on the concentric ring magnets as a function of displacement from equilibrium position for a given current \cite{wang_observation_2018}.

Dissipative forces are modeled phenomenologically with a
viscous damping term, 
\begin{equation}\label{eq:fLoss}
    F_{loss,n} = c \frac{d {u}_{n}}{d t}
\end{equation}
where the damping coefficient $c$ $([c] = \text{Ns/m})$ is determined empirically by minimzing the difference (as a function of $c$) between the simulated and experimental spatial decay of the velocity amplitude envelope of waves traveling through the lattice. A representative example is shown in Fig.~\ref{fig:schematic_photo}(d).
The value of $c$ that minimized the magnitude of the difference  is used in all analytical and numerical modeling.

The coupling force term is defined using the repulsive magnetic force $P(x)$ between neighboring masses, where $P(x)$ is a function of the center-to-center distance $x$ (m) between masses. The measured force-distance relation between neighboring masses is fit with a dipole-dipole approximation given by
\begin{equation}\label{eq:Pdipole}
    P_{dipole} (x) = k_{dipole} x^{-\alpha} + P_{0,dipole}.
\end{equation}
The measured and dipole-dipole fit values are shown in Fig.~\ref{fig:schematic_photo}(c).
If the displacement amplitude of the masses is small relative to the equilibrium distance between adjacent masses,
it is useful to employ a linear approximation using the Taylor expansion of the above expression about the equilibrium displacement
\begin{equation}\label{eq:Plin}
    P_{linear} (x) = k_{linear} x,
\end{equation}
where $k_{linear} =  P_{dipole}'(a)$,
where $a$ is the equilibrium distance between adjacent masses.
Thus the total coupling force on a given mass is calculated using its displacement $u_{n}$, the displacement of the neighboring masses $u_{n\pm1}$ as follows
\begin{equation}\label{eq:Fc}
    F_{mag,n} = P\left(a - u_{n} + u_{n+1} \right) - P\left(a - u_{n-1} + u_{n} \right),
\end{equation}
where $P$ can be given by either $P_{dipole}$ or $P_{linear}$.

\begin{table*}
\caption{\label{tab:paramTable}Lattice Model Parameters}
\begin{tabularx}{\textwidth}{XXXX}
\hline\hline
 \multicolumn{2}{c}{Measured}&\multicolumn{2}{c}{Fit}\\
 Parameter &  Value & Parameter &  Value \\ \hline
 $m$ & 0.0097 [kg] & $c$ & 0.15 [N.s/m] \\
 $a$ & 33.4 [mm] & $k_{linear}$ & 87.03 [N/m] \\
 $f_{mod} = \frac{\omega_{mod}}{2\pi}$ & $\in [1\;\;70]$ [Hz] & $k_{dipole}$ & $9.044\times10^{-7} \text{Nm$^{4}$}$ \\
 $f_{dr} = \frac{\omega_{dr}}{2\pi}$ & $\in [1\;\;40]$ [Hz] & $P_{0,dipole}$ & 0.7047 [N] \\
 ~ & ~ & $A_{mod}$ & $\in [0\;\; 100]$ [N/m]\\
 ~ & ~ & $A_{dr}$ & $\in [0.15\;\;0.4]$ [N] \\ \hline\hline
\end{tabularx}
\end{table*}

Table \ref{tab:paramTable} summarizes the measured and fit parameters used throughout the manuscript.

\section{Wavenumber Bandgaps} \label{sec:wnbg}

\subsection{Theoretical Determination of Wavenumber Bandgaps} \label{sec:disp}

\begin{figure}
    \centering
     \includegraphics[width=\linewidth]{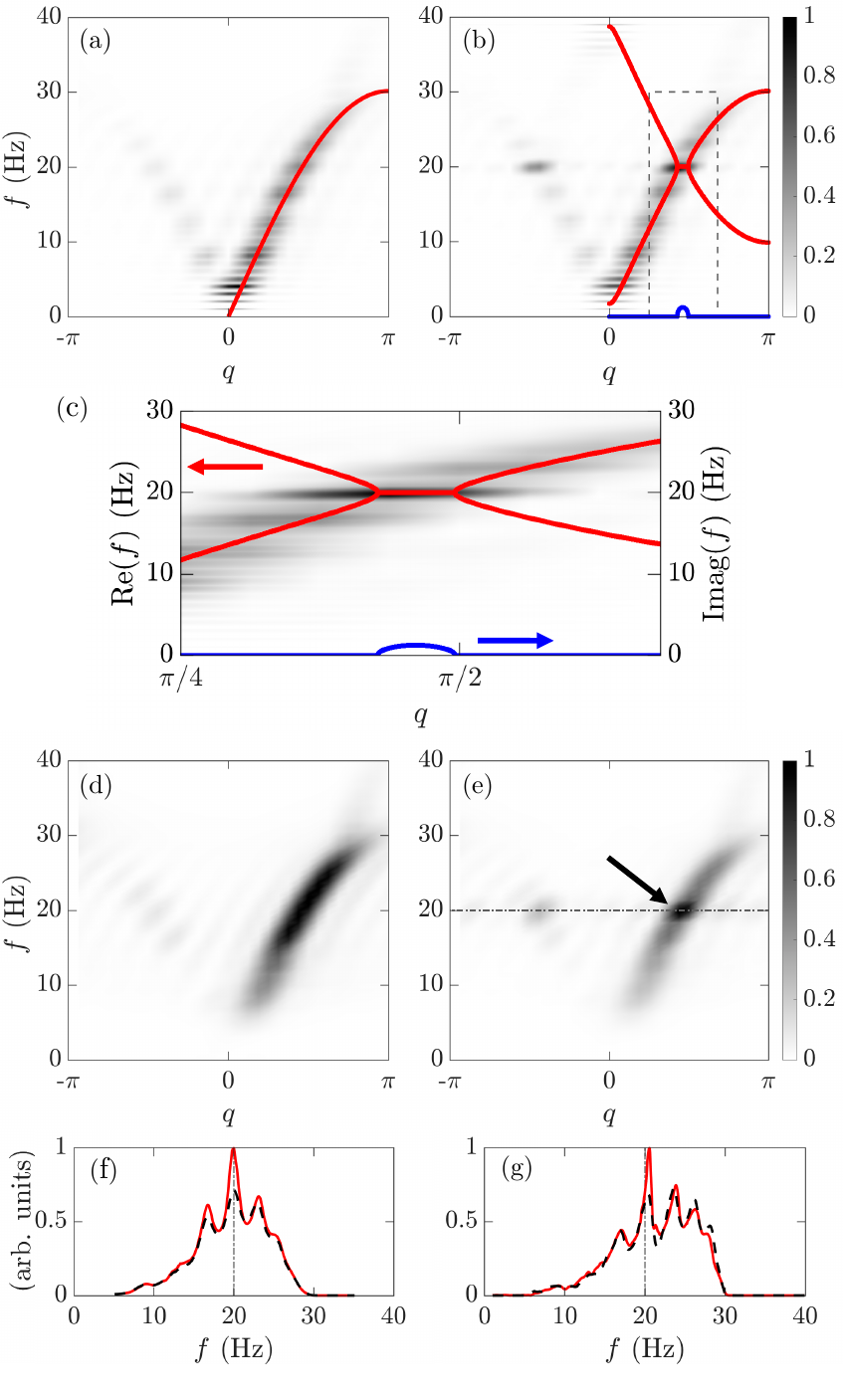}\\
    \caption{(Color online) Dispersion relation and transmission response in unmodulated and modulated lattice at $f_{mod} = 40$ Hz. \textbf{(a)} Numerical dispersion reconstruction for the unmodulated lattice. The analytical predication is shown by the red curve. \textbf{(b)} Numerical dispersion reconstruction with modulation frequency $f_{mod} = 40$ Hz. The real part (red curve) and imaginary part (blue curve) of the analytical approximation is also shown. \textbf{(c)} Expanded view of the bandgap from the gray dashed window in panel (b).
    \textbf{(d)} Experimentally measured dispersion reconstruction for the unmodulated lattice. \textbf{(e)} Experimentally measured dispersion reconstruction with modulation frequency $f_{mod} = 40$ Hz. The frequency $f_{mod}/2 = 20$ Hz is indicated by the gray dash-dotted line. The arrow highlights amplitude peak in dispersion branch. \textbf{(f)}
    Numerically simulated frequency transmission spectra for the unmodulated lattice (black dashed curve) and modulated lattice (red curve) with $f_{mod} = 40$ Hz. The frequency $f_{mod}/2 = 20$~Hz is indicated by the gray dash-dotted line.
   \textbf{(g)} Same as panel (f), but for the experimentally measured frequency transmission spectra.
    }
    \label{fig:dispersion}
\end{figure}

In the small amplitude displacement regime, we observe experimentally the existence of wavenumber bandgaps. In this section, we summarize the linear theory predicting parameter values that lead to the emergence of wavenumber bandgaps in our system. In the limit of small displacements, we employ the linear approximation of the magnetic inter-site-coupling discussed previously. Ignoring damping (which we return to later) and assuming  that
the chain is infinite in length results in the following model
\begin{eqnarray}
    m\frac{d^{2}{u}_{n}}{d t^{2}} + A_{mod} \cos{\left(2\pi f_{mod} t \right)} u_{n}&&\nonumber
    \\+ k_{linear} \left( 2 u_{n} - u_{n-1} - u_{n+1} \right) &&= 0. \label{eq:EOMnoDamp}
\end{eqnarray}
One approximate solution of this equation will be the sum of the incident wave and the scattered fields induced by the 
time modulation \cite{nassar_non-reciprocal_2017},

\begin{eqnarray}\label{eq:planeWave}
    u_{n}(t) = &&U_{-1} e^{i\left(q_{0} n - 2\pi f_{-1} t\right)}  +  U_{0} e^{i\left(q_{0}n - 2\pi f_{0} t\right)}\nonumber\\
    &&+ U_{1} e^{i\left(q_{0} n - 2\pi f_{1} t\right)},  
\end{eqnarray}
where $f_{0}$ is the ordinary frequency of the incident wave with amplitude $U_0$ and $q_{0} = \kappa_{0}a$ is the dimensionless form of the wavenumber. The amplitudes of the scattered fields are $U_{-1}$ and $U_1$ which have frequencies $f_{\pm 1} = f_0 \pm f_{mod}$.
The scattered field is negligible except where it intersects the incident field \cite{nassar_non-reciprocal_2017,wang_observation_2018}. This implies $D \left( f_{0}, q_{0} \right) = D \left( f_{\pm 1}, q_{0} \right) = 0$, 
where
\begin{equation}\label{eq:unmodDispersion}
    D \left( f, q \right) := m (2\pi f)^{2} - 4 k_{linear} \sin^{2}\left( q/2 \right) = 0
\end{equation}
is the usual dispersion relationship in the unmodulated lattice (i.e. $A_{mod} = 0$).
The intersections occur precisely when $f_0(q) = \frac{1}{2} f_{mod}$.
Substituting Eq.~\eqref{eq:planeWave} into Eq.~\eqref{eq:EOMnoDamp} and equating coefficients of the three harmonics leads to a zero determinant condition \cite{nassar_non-reciprocal_2017}, and is given by the expression
\begin{widetext}
\begin{equation}\label{timeModDispersion}
   \hat{D}(f, q) = D \left( f - f_{mod}, q \right)D \left( f, q \right)D \left( f + f_{mod}, q \right) - A_{mod}^2 \left(D \left( f - f_{mod}, q \right) - D \left( f + f_{mod}, q \right) \right) = 0.
\end{equation}
\end{widetext}
This condition is a modified dispersion relation in the neighborhood of the intersections of $D \left( f_{\pm 1}, q_{0} \right) = 0$ and $D \left( f_{0}, q_{0} \right)=0$. Values of $q$ that lead to solutions of Eq.~\eqref{timeModDispersion} with complex valued $f$ makeup the so-called wavenumber bandgaps in the band structure and correspond to a parametrically amplified standing wave with growth rate given by the imaginary part of $f$ \cite{trainiti_time-periodic_2019,lee_parametric_2021}.
The analytical dispersion relations
for the unmodulated
($D(f,q)=0$) and modulated ($\hat{D}(f,q)=0$)
lattices  are shown in Fig.~\ref{fig:dispersion}(a) and (b), respectively, for parameters $f_{mod} = 40$ Hz and $A_{mod} = 37.5$ N/m. As predicted by the intersection of the incident and scattered fields, the wavenumber bandgap opens at $f = f_{mod}/2$, as seen in Fig.~\ref{fig:dispersion}(c).

To verify the dispersion calculations, we simulate Eq.~\eqref{eq:EOM} using the same modulation parameters for the grounding stiffness $k_{g}(t)$ and mass $m$ that were used to compute
the dispersion relationships, but we use the nonlinear repulsive force ($P_{dipole}$ instead of $P_{linear}$), and we include viscous damping $c$ (see Table \ref{tab:paramTable} for the specific values used). The simulation is solved repeatedly for monochromatic, six-cycle sine bursts from $f_{dr} = 1$ to 40~Hz (in 1 Hz increments) with driving amplitude $A_{dr} = 0.38$~N. The numerical dispersion
relationship is obtained by computing the two-dimensional Fourier transform (2DFFT) of the velocity components of the numerical solutions.
Color intensity corresponds to the normalized spectral energy density (a composite of all driving frequencies) of the unmodulated ($A_{mod}=0$ N/m)
and modulated ($A_{mod} = 37.5$ N/m) lattices in Fig.~\ref{fig:dispersion}(a) and (b), respectively. This method is similar to a spectral energy density method frequently employed in photonic and phononic systems \cite{vila_bloch-based_2017,nassar_non-reciprocal_2017,wang_observation_2018}. The numerical dispersion relation accounts for nonlinearity, gain, losses, and finite effects
but still agrees well with the linear theory based on the infinite losses lattice
with only three fields used to determine the dispersion relationship.

\subsection{Experimental Observation of Wavenumber Bandgaps}

To reconstruct the dispersion relation of acoustic waves propagating through the experimental lattice, we measure the velocity of each mass using a laser Doppler vibrometer (LDV, Polytec CLV-2534). We then construct a full space-time-resolved transient velocity response of the lattice, excited by single-frequency, six-cycle sine wave bursts from quiescent initial conditions (the same method used in the numerical simulations). In the finite experimental and simulated lattices, the finite-cycle burst and measurement time window are chosen so that reflections off the boundary are not captured. Using the velocity field measurements from an appropriate range of burst frequencies, a composite of spectral energy densities of the two-dimensional velocity fields yields a reconstruction of the dispersion relation. Fig.~\ref{fig:dispersion}(a), (b), (d), and (e) show the comparison of measured dispersion against the numerical simulation overlaid with the analytical predictions.

The measured dispersion reconstruction in Fig.~\ref{fig:dispersion}(d) for the lattice without external modulation exhibits the expected behavior of a monatomic lattice, i.e., a single acoustic branch terminating at the edge of the Brillouin zone.
Both positive and negative wavenumbers are plotted, which may be interpreted here as forward and backward propagating waves, respectively.

The measurements are then repeated with an extrinsic temporal grounding stiffness modulation applied to the lattice via the electromagnetic coils. The effective grounding stiffness of the masses are modulated harmonically at $f_{mod} = 40$ Hz. Dissipation in the experimental apparatus makes detection of small wavenumber signals difficult; therefore, the modulation frequency is selected so that the salient features of the wavenumber bandgap, which occur at $f_{mod}/2$, lie on a clear section of the dispersion branch. The numerical and experimental dispersion reconstructions shown in Fig.~\ref{fig:dispersion}(b) and (e), respectively, exhibit a strong peak (i.e. darker regions in the spectral energy density), at $f_{mod}/2 = 20$ Hz on the dispersion branch. In the numerical simulation, this peak aligns with the analytical prediction of the wavenumber bandgap, and the experimental peak is highlighted by an arrow and seen to align with $f_{mod}/2 = 20$ Hz. This increased amplitude response is consistent with the expected parametric amplification, associated with the complex frequency inside the wavenumber bandgap, and is in line with previous results in the literature \cite{trainiti_time-periodic_2019,lee_parametric_2021}. Compared to the unmodulated lattice, the dispersion branch of the modulated lattice is largely unchanged, except for the small neighborhood of frequencies around $f_{mod}/2 = 20$ Hz.

Both the experimental and numerical results show good agreement in the presence of time modulation, and the location of the wavenumber bandgap is predicted accurately by the analytical model. Moreover, an additional amplitude peak is observed at $f_{mod}/2$ and the negative of the wavenumber corresponding to the bandgap, where the negative wavenumber may be interpreted as backward propagating waves (Fig.~\ref{fig:dispersion}(b), (d)). Such behavior is consistent both with the predicted parametrically amplified standing wave solution that occurs within the wavenumber bandgap 
and with previous experimental work that has shown evidence of the same effect via the conversion of broadband signals into narrowband reflections \cite{trainiti_time-periodic_2019}. 

In addition to exploration of dispersion, we examine the transmission spectrum in a two-port configuration for the effectively steady state response to harmonic input. The continuous sine sweep from $f_{dr} = 1$ to 40 Hz reference signal of a lock-in amplifier (LIA, Stanford Research SR860) is applied to the input mass $n = 2$, and the output velocity at mass $n = 11$ is measured using the LDV. The output is filtered against the input by the LIA. Experimental parameters are identical to the dispersion reconstruction, with $f_{mod} = 40$ Hz, $A_{mod} = 37.5$ N/m, and $A_{dr} = 0.38$ N. This is also reproduced in numerical simulation. The resulting frequency spectrum demonstrates clearly that the extrinsic time-periodic modulation induces amplification of signals at half the modulation frequency, in both experiment and simulation, see Fig.~\ref{fig:dispersion}(f) and (g), respectively. The relatively narrowband amplification provides further evidence that the dispersion reconstruction accurately depicts the localized nature of the wavenumber bandgap and its amplifying effect on incident signals.


\subsection{Parametric Investigation of Stability}

A more complete analysis of stability can be achieved by considering more than the three modes included in Eq.~\eqref{eq:planeWave} that result in a complex-valued $f$.
Moreover, the inclusion of damping also has a non-trivial effect on the stability properties.
Therefore, we conduct a stability analysis on the linearized equations of motion of an infinite mass-spring chain with damping,
\begin{eqnarray}\label{eq:EOMDamp}
    m\frac{d^{2}{u}_{n}}{d t^{2}} + c \frac{d{u}_{n}}{d t} + A_{mod} \cos{\left(2\pi f_{mod} t \right)} u_{n}&&\nonumber\\
    + k_{linear} \left( 2 u_{n} - u_{n-1} - u_{n+1} \right) &&= 0.
\end{eqnarray}
In particular,
we make use of discrete Fourier transform,
\begin{equation}\label{eq:DFT}
    \hat{u}(q,t) = \sum_{n\in\mathbb{Z}} u_{n}(t) e^{iqn},
\end{equation}
to cast Eq.~\eqref{eq:EOMDamp} in Fourier space,
\begin{eqnarray}\label{eq:DFTEOM5}
    \partial_{t}^{2}\hat{u}(q,t) + \frac{c}{m }\partial_{t}\hat{u}(q,t)&&\nonumber\\
    + \left[ (2\pi f(q))^{2} + \frac{A_{mod}}{ m} \cos(2\pi f_{mod} t) \right] \hat{u}(q,t) &&= 0,
\end{eqnarray}
where $D(f(q),q)=0$, that is $f(q)$
satisfies the dispersion relation
in the unmodulated lattice. Equation~\eqref{eq:DFTEOM5} is a Mathieu equation, which includes a linear viscous damping term \cite{richards_stability_1976,xie_dynamic_2006,kovacic_mathieus_2018,kutz_advanced_2020}. We apply the transformation
\begin{equation}\label{eq:MathieuTansformation}
    \hat{v}(q,t) = e^{(c/m)/2 t}\hat{u}(q,t),
\end{equation}
which reduces Equation~\eqref{eq:DFTEOM5} to the undamped form of Mathieu equation,
\begin{widetext}
\begin{equation}\label{eq:MathieuTansformed}
    \partial_{t}^{2}\hat{v}(q,t) + \left[ \left((2\pi f(q))^{2} - \frac{(c/m)^{2}}{4} \right) + \frac{A_{mod}}{ m} \cos(2\pi f_{mod} t) \right] \hat{v}(q,t) = 0,
\end{equation}
\end{widetext}
which has the general solution
\begin{equation}
\hat{v}(q,t) = C_1 e^{\mu_+(q) t} p_+(t) + C_2 e^{\mu_-(q) t} p_-(t),
\end{equation}
where $\mu_{\pm}(q)$ are the Floquet exponents associated to wavenumber $q$
and $p_{\pm}(t)$ is a
$1 /f_{mod}$
periodic function. The Floquet
multipliers, defined as
$\sigma_\pm(q) = e^{\mu_\pm(q) /f_{mod} }$,
are the eigenvalues of the fundamental solution matrix evaluated at
$t = 1 / f_{mod}.$
In order for $\hat{u}(q,t)$ to be stable, we must have that $\text{Re}{\left(\mu_\pm(q)\right)} \leq \frac{c/m}{2}$.
Note, if one makes the simplifying assumptions
that $c = 0$ and
$i \mu(q) = 2\pi f(q)$,
and only the $j=-1,0,1$ modes
of the Fourier series expansion of
$p(t) = \sum_{j\in \mathbb{Z}} a_j e^{i j 2\pi f_{mod} t}$
are kept,
one obtains the approximate solution Eq.~\eqref{eq:planeWave} discussed in Sec.~\ref{sec:disp}.

To construct a stability diagram in the $[A_{mod},f_{mod}]$ parameter plane, the
Floquet exponent
pairs $\mu_\pm(q)$
are computed as a function of $q$ for a fixed parameter
set. The wavenumbers for a finite lattice of length $N-1$ with boundaries fixed to zero are given by $q_j = \pi j / N $.
Thus, for a particular parameter set, the dynamics will be stable if 
\begin{equation} \label{eq:fmstab}
\max_{1\leq j \leq N-1} \text{Re}\left(\mu_\pm(q_{j})\right) \leq \frac{c/m}{2} .
\end{equation} 

\begin{figure*}
\centering
    \includegraphics[width=.75\linewidth]{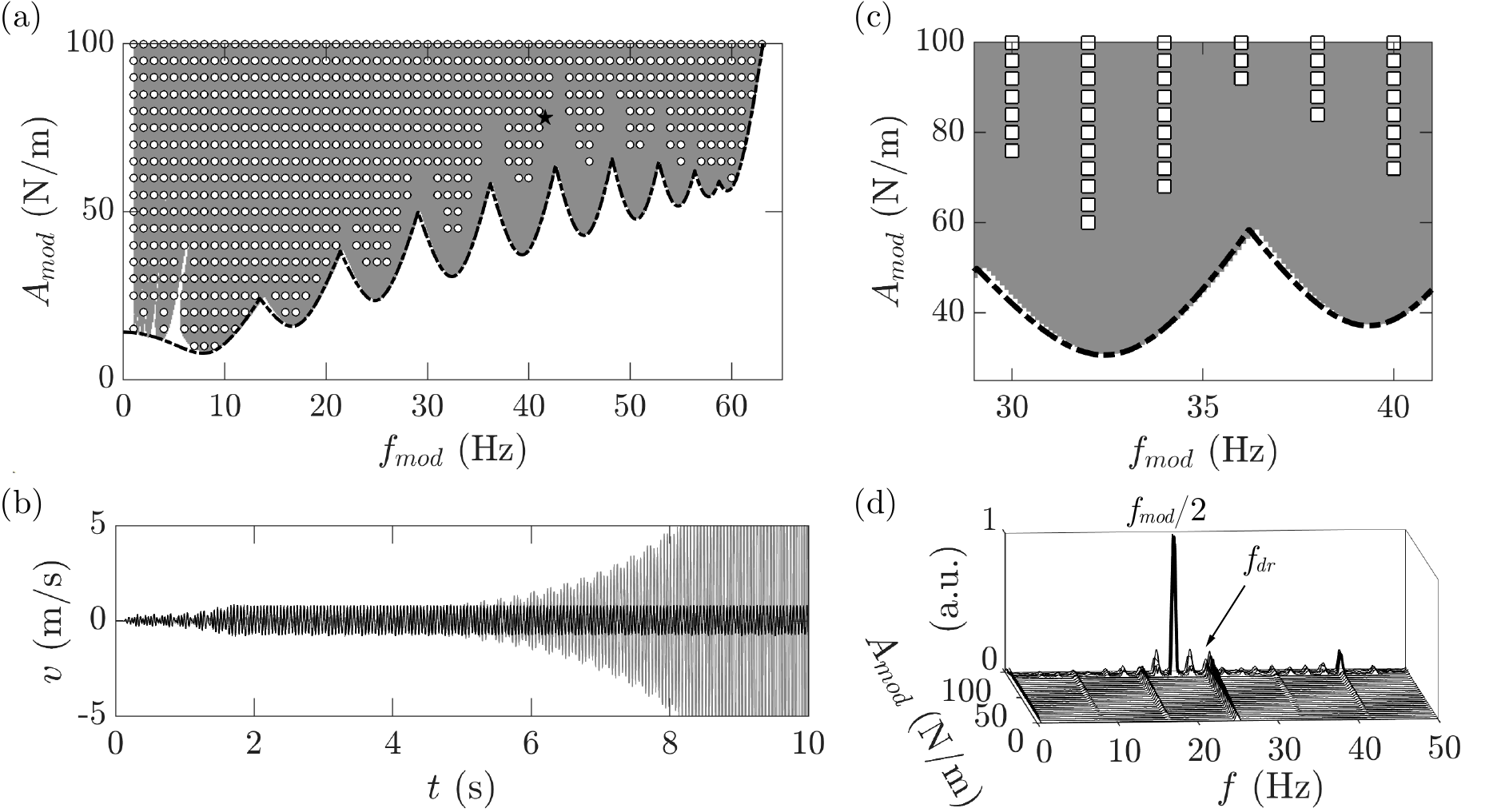}\\
    \caption{Stability of modulation parameters. \textbf{(a)} Shaded region indicates unstable solutions for modulation parameter combinations as determined by the Floquet exponent condition Eq.~\eqref{eq:fmstab}. The black dashed curve shows the analytical prediction of the stability boundary based on condition Eq.~\eqref{eq:stab}.  The white circles indicate parameter combinations for which the fully nonlinear simulation exhibits a non-decaying, modulation-driven response to an initial impulse. The black star indicates the parameters shown in panel (b). \textbf{(b)} Numerically simulated velocity output time series for the parameters indicated by the black star in panel (a). The fully nonlinear simulation (black line) has a bounded response, while the linearized simulation (gray line) exhibits exponential growth. \textbf{(c)} Experimental non-decaying parameter combinations (white squares) overlaid on the Floquet exponent condition Eq.~\eqref{eq:fmstab} (shadded region) and the  Mathieu instability condition Eq.~\eqref{eq:stab} (black dashed curve). \textbf{(d)} The frequency response of the experimental lattice for fixed amplitude harmonic driving and increasing harmonic modulation. At a critical amplitude, the lattice transitions from a driving-dominated response to a high amplitude, modulation-dominated response.}
    \label{fig:stabilityDiagram}
\end{figure*}

Regions in parameter space that are unstable according to the above condition
are shown as the gray shaded regions of Fig.~\ref{fig:stabilityDiagram}(a).

The stability regions of the Mathieu equation can also be approximated analytically. 
The standard form of the Mathieu equation is
$\Ddot{x}(t) + \gamma \Dot{x}(t) + \left( \delta + \epsilon \cos (t) \right) x(t) = 0$
where $\gamma = \frac{c}{m2\pi f_{mod}}$, $\delta = \left( \frac{f(q)}{f_{mod}} \right)^{2}$, and $\epsilon = \frac{A_{mod}}{(2\pi f_{mod})^{2} m}$. In the absence of damping, regions of instability in the $[\delta,\epsilon]$ parameter plane emerge at the values  $\delta_j = \frac{j^2}{4} $, where $j$ is an integer \cite{nayfeh_nonlinear_2004}.
Since $f(q)$ increases monotonically from zero, a condition for stability can be obtained by considering the instability tongue associated to $j=1$, namely that $\left( \frac{f(q)}{ f_{mod}} \right)^{2} < 1/4$, which recovers the already obtained result that instability is induced by the intersection of dispersion curves when  $f(q) = f_{mod}/2$.
This condition reproduces the incident-scattered field dispersion intersection condition discussed in Section \ref{sec:disp}.

The transition curves of the the stability regions of the Mathieu equation with damping can be found via perturbation analysis~\cite{kovacic_mathieus_2018,nayfeh_nonlinear_2004},
\begin{equation} \label{eq:stab_scaled}
    \delta = \frac{1}{4} \pm \frac{\sqrt{\epsilon^2 - \gamma^2}}{2},
\end{equation}
which is valid for small $\epsilon$. In terms of the original system parameters, and 
taking into account a finite lattice, in which the wavenumbers become $q_j = \pi j / N $,
Eq.~\eqref{eq:stab_scaled} implies the following condition for stability
\begin{widetext}
 \begin{equation} \label{eq:stab}
\max_{1\leq j \leq N-1} \left| \left(\frac{f(q_{j})}{f_{mod}}\right)^{2} \pm \frac{1}{2} \sqrt{ \left(\frac{A_{mod}}{ (2\pi f_{mod})^2 m}\right)^{2} - \left(\frac{c}{m2\pi f_{mod}}\right)^{2}} \right | < \frac{1}{4}.
\end{equation}
\end{widetext}
The black line of Fig.~\ref{fig:stabilityDiagram}(a) shows the transition curves of the regions of instability based on the analytical approximation Eq.~\eqref{eq:stab}.

In order to explore the validity of the linear theoretical stability predictions, we study both the numerically simulated and experimentally measured response of the lattice. The existence of unstable, exponentially growing solutions from the linear model implies large-amplitude displacements, and indeed large-amplitude displacements (relative to the equilibrium spacing $a$) are experimentally observed.
To illustrate the difference between the responses, we simulate the response of the lattice both with the linearized and the nonlinear repulsive force ($P_{linear}$ and $P_{dipole}$ respectively)
using the measured and fit parameters matching the experimental setup (see Table~\ref{tab:paramTable}). An unstable set of modulation parameters, as predicted by the linear theory (in particular, $f_{mod} = 41.6$ Hz, $A_{mod} = 78$ N/m, see black star in Fig.~\ref{fig:stabilityDiagram}(a)), is applied to the lattice with no input drive ($A_{dr} = 0$). The simulation is initiated with quiescent conditions except for an initial velocity at the driving mass ($n=2$). It is observed that while the response of the linear simulation grows exponentially, the nonlinear simulation reaches an oscillatory steady state, sustained by the grounding stiffness modulation. This is illustrated by the velocity responses of the output mass ($n = 11$) in Fig.~\ref{fig:stabilityDiagram}(b), with the linear simulation in gray and nonlinear simulation in black. Conversely, for parameter values where the linear theory predicts stability,
the responses of both the linear and nonlinear simulations decay with time due to damping. Thus, as a proxy for the theoretical linear instability, we search the full modulation parameter space for any response that does not decay with time, what we will refer to as non-decaying responses, from the nonlinear simulation or experimental lattice.

The modulation parameters that lead to a non-decaying response for the numerical simulation are denoted in Fig.~\ref{fig:stabilityDiagram}(a) by white markers. The region of modulation parameters that lead to a non-decaying response in the numerical simulation with nonlinear interaction included exhibits good quantitative agreement with the unstable region predicted through both the Floquet theory (Eq.~\eqref{eq:fmstab}) and damped Mathieu (Eq.~\eqref{eq:stab}) stability conditions. We perform the same procedure for the experimental setup over a subset of the modulation parameter space, exciting the input mass ($n=2$) with an impulse and observing decaying or non-decaying responses. The experimental non-decaying region (Fig.~\ref{fig:stabilityDiagram} (c)) shows similarly good agreement with the linear theoretical predictions.

Despite linearization and, in the case of the Mathieu condition, a perturbation method approximate solution, both the Floquet and Mathieu stability conditions accurately predict the ranges of modulation parameters for which a modulation-driven response is observed in the experiment. This suggests that the onset of such modulation-driven response can be predicted, to a degree, by the approximate linear dynamics of the lattice. 

When the lattice is driven harmonically, and the modulation amplitude is incrementally increased, we observe a transition from a driving-dominated to a modulation-dominated response, as shown in Fig.~\ref{fig:stabilityDiagram}(d) for the measured steady state output (mass $n=2$) velocity. The frequency response shows a sharp transition in dominant frequency component (from driving frequency to half the modulation frequency) and large increase in amplitude. The time series and frequency responses (Fig.~\ref{fig:stabilityDiagram}(b) and (d)) suggest the nonlinear interaction force bounds the response, even for linear unstable modulation parameters. This bifurcation and nonlinear bounded states are considered in more detail in the following section.


\begin{figure*}
    \centering
     \includegraphics[width=0.8\linewidth]{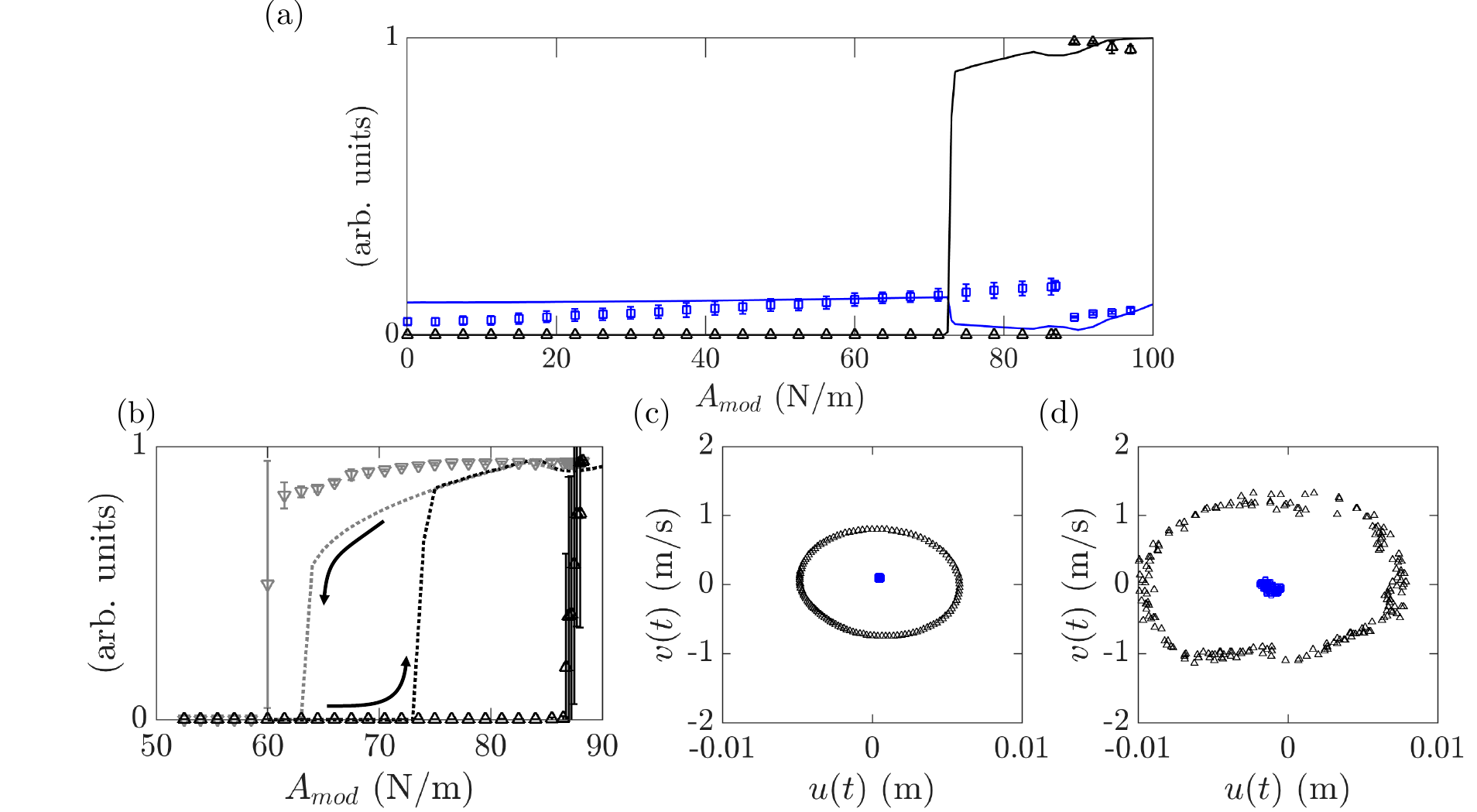}
     
    \caption{(Color online) Nonlinear lattice dynamics \textbf{(a)} Fourier amplitudes for $f_{dr}$ (blue/squares) and $f_{mod}/2$ (black/triangles) versus modulation amplitude for experimental (markers) and numerical simulation (lines). Note that error bars are also shown.
    \textbf{(b)} Hysteresis of $f_{mod}/2$ Fourier component around mode transition, with slowly increasing (black triangles) and decreasing (gray upside down triangles) modulation amplitude. Numerical simulation shown in dashed lines. 
    \textbf{(c)} Poincar\'e section of the output response of the numerical simulation, with sampling period $T = 1/f_{dr}$. 
    Pre- and post- forward sweep transition modulation amplitude responses are shown as blue squares and black triangles, respectively. \textbf{(d)} Same as panel (c) for the experiment.}
    \label{fig:nonlinearLattice}
\end{figure*}

\section{Nonlinear Lattice Dynamics}\label{sec:nonlinear}

We now further investigate the interplay of the nonlinearity of the system with the extrinsic time modulation. In the previous section, we observed that the nonlinearity has a stabilizing effect, leading to bounded steady-states rather than unbounded growth as the linear theory predicts.
We will examine the structure of these nonlinear steady-states and explore the associated bifurcations as a system parameter is varied.

We start by conducting a parametric sweep of the modulation amplitude $A_{mod}$, both experimentally and numerically. This is a natural
parameter to consider for bifurcation studies, since it is expected that larger values of $A_{mod}$ will lead to nonlinear effects.
For each value of $A_{mod}$, the lattice is driven by a harmonic input at one end ($n=2$), and the output signal is measured at the opposite end ($n=11$). The velocity response is allowed to reach steady-state and the amplitude of the response is recorded. In particular, the magnitude of the temporal Fourier coefficient associated to the drive frequency $f_{dr}$ and half the modulation frequency $f_{mod}/2$
are recorded. This will indicate if the observed dynamics is due primarily to the drive (i.e. larger Fourier amplitude at $f=f_{dr}$) or the time modulation 
(i.e. larger Fourier amplitude at $f=f_{mod}/2$). The modulation amplitude is increased by increment $\Delta A_{mod}$, and the response is again allowed to reach steady-state and is recorded. These steps are repeated until the maximum modulation amplitude is reached. We call this procedure the "forward sweep". The "backward sweep" procedure is similar, where $A_{mod}$ is decreased rather than increased. This process is carried out numerically
and experimentally. In particular, for the numerical results, Eq.~\eqref{eq:EOM} is simulated with the parameters specified in Table~\ref{tab:paramTable} and with
drive frequency $f_{dr} = 25$ Hz and drive amplitude $A_{dr} = 0.15$ N. The modulation frequency $f_{mod} = 41.6$ Hz is chosen to be incommensurate with the driving frequency $f_{dr}$ in order to demonstrate general quasiperiodic behavior. The range of modulation amplitudes considered is $A_{mod} \in [0,100]$ N/m where increments of $\Delta A_{mod} = 2$N/m are used
in the sweeps. Experimental forward and backward sweep measurements are repeated 4 times with identical driving and modulation parameters, with the exception that the step size is $\Delta A_{mod} \approx 3.75$.

Fig.~\ref{fig:nonlinearLattice} summarizes the results of the modulation amplitude sweeps. 
In Fig.~\ref{fig:nonlinearLattice}(a) the results of the forward-sweep are shown. The Fourier amplitudes associated to the drive frequency $f=f_{dr}$ are shown in blue (squares/lines) and the Fourier amplitudes of the modulation $f=f_{mod}/2$ are shown in black (triangles/dash-dot lines). Error bars show standard deviation for experimental measurements.
For small modulation amplitudes, the response is dominated by drive dynamics (notice the magnitude of the blue markers compared to the black markers in Fig.~\ref{fig:nonlinearLattice}(a)).
At a critical modulation amplitude, the output response transitions sharply from the small displacement, driving signal-dominated regime to a large displacement, modulation-dominated regime. This transition occurs at approximately $A_{mod}=90$N/m in the experiments and $A_{mod}=73$ N/m in simulation.
 
Fig.~\ref{fig:nonlinearLattice}(b) shows a comparison of the Fourier amplitude of $f=f_{mod}/2$  of the forward sweep (black) and the backward sweep (gray) near the transition point.
Hysteric behavior is observed both in the experiment (markers) and simulation (dashed-lines), with the region of bi-stability
being slightly larger in the experiment. The Poincar\'e sections of the output response give a better sense of the structure of the steady-state.
In particular the output position $u$ (which is found experimentally by integrating the velocity measured by the LDV) and velocity $v=\frac{du}{dt}$ are plotted in the $(u,v)$ phase plane with a sampling period of $T = 1/f_{dr}$. Both numerical simulation and experimental Poincar\'e sections are sampled at modulation amplitudes at least one $\Delta A_{mod}$ smaller and larger than their respective forward sweep transition points. Fig.~\ref{fig:nonlinearLattice}(c) shows the numerically simulated Poincar\'e section at $A_{mod} = 71$ N/m  (blue squares) and $A_{mod} = 75$ N/m (black triangles) and Fig.~\ref{fig:nonlinearLattice}(d) shows the experimentally measured Poincar\'e section at $A_{mod} \approx 80$ N/m  (blue squares) and $A_{mod} = 93$ N/m (black triangles).
Before the transition, the plot of the Poincar\'e sections reveals a single point, indicating the solution is time-periodic. After the transition, the Poincar\'e sections
form an invariant curve in the phase plane, indicating the solution is temporally quasi-periodic.
The Poincar\'e sections confirm what is already suggested in Fig.~\ref{fig:nonlinearLattice}(a). Namely, there is a single dominant frequency in the response before the transition (time-periodic response)
and there are two non-negligible incommensurate frequencies after the transition (time-quasi-periodic response).

In summary, the nonlinearity of the system along with the time-modulation allows for the creation of stable, large-amplitude time-quasi-periodic
solutions that can co-exist with stable, small-amplitude time-periodic ones. This was confirmed both numerically and experimentally,
with the experiments showing good qualitative agreement with the numerics. While additional tuning of the parameters could yield
better quantitative agreement, the primary features of nonlinearity and time modulation are captured well by our model with predetermined parameter values.

\section{Summary and Conclusions}\label{sec:theend}

We studied the response of a linear and nonlinear discrete, phononic lattice, consisting of magnetic particles controlled by electromagnetic coils. We excited the lattice at one end and imparted external stiffness modulation at each particle site. In the linear regime, we experimentally reconstructed the dispersion relation of a chain with modulated grounding stiffness, demonstrating the opening of wavenumber bandgaps. For larger modulation amplitudes, the nonlinearity of the coupling force between masses admits bounded solutions that would otherwise not be present in a linear system, where the parametric amplification characteristic of this form of extrinsic modulation induces exponential growth. Our analysis offers validation of the linear dynamics that produce the unique emergent dispersive properties of time-modulated systems, while demonstrating how nonlinearity provides  additional flexibility in the design and study of wave propagation time-modulated systems. The findings offer insights on methods to control the propagation of acoustic waves in nonlinear, active systems. Implementing such solutions in small scale devices holds promise for applications in sensing and signal processing, offering frequency agile solutions for tunable filters, delay lines and signal conversion. Such nonlinear phenomena could also be used to compensate losses and dissipation, thereby allowing the miniaturization of components and the addition of on-chip functionalities.

\begin{acknowledgements}
This material is based upon work supported by the US National Science Foundation under Grant Nos. DGE‐1745301 (B.L.K.), EFRI-1741565 (C.D.) and DMS-2107945 (C.C.). We thank Prof. Bumki Min and Dr. Jagang Park for helpful discussions.\\
\end{acknowledgements}

\nocite{*}

\bibliography{main}

\end{document}